\documentclass[sigconf]{acmart}

\renewcommand\footnotetextcopyrightpermission[1]{} 
\usepackage{tabularx}

\newcolumntype{C}{>{\centering\arraybackslash}X}

\usepackage{graphicx} 
\graphicspath{ {./supplements/} }

\AtBeginDocument{%
  \providecommand\BibTeX{{%
    \normalfont B\kern-0.5em{\scshape i\kern-0.25em b}\kern-0.8em\TeX}}}

\setcopyright{none} 
\makeatletter
\def\@copyrightspace{\relax}
\makeatother

\begin{document}

\title{A Mining Software Repository Extended Cookbook: Lessons learned from a literature review}


\author{Daniel D. R. Barros}
\email{daniel.barros@ufabc.edu.br}
\affiliation{%
  \institution{Federal University of ABC}
  \streetaddress{Av. dos Estados, 5001}
  \city{Santo André}
  \state{São Paulo}
  \country{Brazil}
  \postcode{09210-580}
}

\author{Flávio Horita}
\email{flavio.horita@ufabc.edu.br}
\affiliation{%
  \institution{Federal University of ABC}
  \streetaddress{Av. dos Estados, 5001}
  \city{Santo André}
  \state{São Paulo}
  \country{Brazil}
  \postcode{09210-580}
}

\author{Kanan C. Silva}
\email{kanan.silva@ufabc.edu.br}
\affiliation{%
  \institution{Federal University of ABC}
  \streetaddress{Av. dos Estados, 5001}
  \city{Santo André}
  \state{São Paulo}
  \country{Brazil}
  \postcode{09210-580}
}

\author{Igor S. Wiese}
\email{igor@utfpr.edu.br}
\affiliation{%
  \institution{Federal University of Technology - Parana (UTFPR)}
  \streetaddress{R. Rosalina Maria Ferreira, 1233}
  \city{Campo Mourão}
  \state{Paraná}
  \country{Brazil}
  \postcode{87301-899}
}

\renewcommand{\shortauthors}{Barros et al.}
\fancyhead{}
\renewcommand{\headrulewidth}{0pt}
\begin{abstract}
The main purpose of Mining Software Repositories (MSR) is to discover the latest enhancements and provide an insight into how to make improvements in a software project. In light of it, this paper updates the MSR findings of \textit{the original MSR Cookbook}, by first conducting a systematic mapping study to elicit and analyze the state-of-the-art, and then proposing an extended version of the Cookbook. This extended Cookbook was built on four high-level themes, which were derived from the analysis of a list of 112 selected studies. Hence, it was used to consolidate the extended Cookbook as a contribution to practice and research in the following areas by: 1) including studies published in all available and relevant publication venues; 2) including and updating recommendations in all four high-level themes, with an increase of 84\% in comments in this study when compared with \textit{the original MSR Cookbook}; 3) summarizing the tools employed for each high-level theme; and 4) providing lessons learned for future studies. Thus, the extended Cookbook examined in this work can support new research projects, as upgraded recommendations and the lessons learned are available with the aid of samples and tools.
\end{abstract}

\maketitle

\vspace{-1mm}

\section{Introduction} \label{sec:introduction}
Software repositories such as version control systems 
contain a huge amount of available data that have given rise to an extensive research endeavor in the field of Mining Software Repositories (MSR) \cite{Poncin.2011}. In light of this, the challenging task of software engineering can be assisted by extracting knowledge during the software development cycles \cite{Xie.2009}; in particular, using software repositories as the main information source along with mining techniques to achieve improvements \cite{Akbar.2020, Paixao.2020}. Here, MSR contributes to discovering enhancements and providing an insight into how to make improvements in a software project. MSR research is already renowned in academia \cite{Hemmati.2013, Luzgin2020} and industry \cite{Hassan.2006, Tripathi.2015} and has become a key topic in important publication venues and research projects \cite{Guemes.2018}. 

From the time when it began in 2004, the MSR field has been represented at the MSR Conference (MSRConf) \footnote{\url{http://www.msrconf.org/}}, this is an important event that seeks to advance MSR science and practices. Since then, MSR research has been growing with the advance of technology and research studies that have provided new techniques, mapping, trends, and a set of best practices. Some mapping studies and surveys have also focused on MSR \cite{Chaturvedi2013, Farias.2016}. Among them, \textit{the original MSR Cookbook} \cite{Hemmati.2013} has been a seminal paper in which several authors have surveyed MSRConf between 2004 and 2012 by formulating a set of best practices and making a number of recommendations. Looking ahead, the current expansion of the MSR field, suggests there are still many opportunities to be addressed.

The challenges are twofold. First, owing to the growing interest in the practice and research on MSR, the literature reviews tend to be out-of-date, including the main reference-point discussed here, \textit{the original MSR Cookbook}, and the mapped works between 2004 and 2012. Second, the previous MSR literature reviews \cite{Farias.2016, Chaturvedi2013, Cosentino.2017, Bavota2016} did not cover the increasing number of conferences and workshops 
that are currently being provided by the MSR field. 

In addressing these challenges, this study seeks to update the MSR research by first carrying out a systematic mapping study to determine and analyze the current state-of-the-art, and then proposing an extended Cookbook. This extended Cookbook focuses on updating the recommendations of the four high-level themes that can be found in \textit{the original MSR Cookbook}, as well as outlining the tools, experiences and achievements listed in the studies of the lessons learned. When undertaking it, this paper set out with 237 studies from the digital libraries available and, after applying the selection criteria, 112 papers were chosen. Out of these studies, 492 comments (notes on the analyzed papers) were categorized in accordance with their subject-areas, and summarized on the basis of the recommendations. As a result, there was an 84\% increase of comments between this study and \textit{the original MSR Cookbook}, several new recommendations were made and some were found to be no longer valid. As well as confirming the interest of the community and the extent to which MSR research had spread, the results also showed an increase in the number of papers that apply mining techniques (e.g., collinearity, text mining, classification, and prediction), together with recently published tools designed to encourage new experiments and further analysis.

Within the context of the research conducted by Mendes et al. \cite{Mendes2020}, this study seeks to make a contribution to the MSR field by updating the state-of-the-art of \textit{the original MSR Cookbook}. In addition, it explains how this topic has evolved in recent years, or in other words, how the findings have changed over time and the reasons why this study was conducted. Finally, on the basis of an analysis of new data, this research investigates further outcomes and expands \textit{the original MSR Cookbook}.


\section{Related works}  \label{sec:relatedwork}
Hemmati et al. \cite{Hemmati.2013} have classified a list of guidelines based on the period 2004 - 2013, which includes papers published at MSRConf. These guidelines, called \textit{the original MSR Cookbook}, were a compilation of comments from papers that were categorized through an open coding approach \cite{Miles.1994}. The compilation of the comments resulted in a list of recommendations that formed the stages of a typical MSR process as follows: i) to start with, data extraction and preprocessing; ii) the application of mining techniques; iii) a performance analysis of data and iv) an interpretation of the results; and finally v) data sharing.

\begin{figure*}[h]
    \centering
    \includegraphics[width=0.65\textwidth]{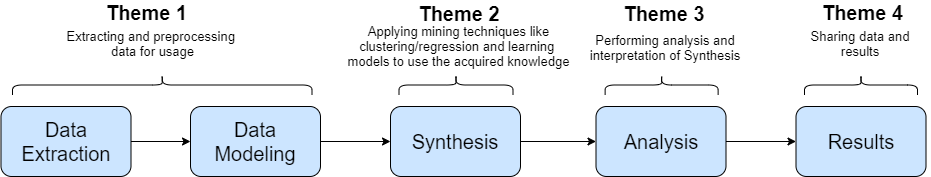}
    \caption{High-level Themes}
    \label{fig:HLThemes}
\end{figure*}

Hence, one of the main research contributions of \textit{the original MSR Cookbook} was that it established four high-level themes inspired by the selected papers: \textit{Theme 1: Data acquisition and preparation}, \textit{Theme 2: Synthesis}, \textit{Theme 3: Analysis}, and \textit{Theme 4: Sharing and replication}. \textbf{Theme 1} focuses on how and what should be preprocessed and extracted from software artifacts, in particular, those collected from software repositories. \textbf{Theme 2} is concerned with applying mining algorithms such as classification, collinearity, and text mining, \textbf{Theme 3} conducts a statistical analysis and makes an evaluation of Theme 2. Finally, \textbf{Theme 4} is concerned with sharing data and results. 
Figure \ref{fig:HLThemes} shows these themes and their features. 

There are some systematic mapping studies on MSR. Farias et al. \cite{Farias.2016} conducted a study based on five editions of MSRConf between 2010 and 2015, and obtained evidence of software analysis, data sources, evaluation methods, tools, and how the MSR area is improving. 
Chaturvedi et al. \cite{Chaturvedi2013} studied the tools employed in software repositories by carrying out a systematic mapping from the MSRConf papers published between 2007 and 2012. Unlike previous studies, this work supplies the gap for researching in new venues, including of course MSR as one of them, as the data extraction considered all the digital libraries feasible and available.

There are also related works 
that investigated MSR by only concentrating on one coding platform. Consentino et al. \cite{Cosentino.2017} mapped studies in social coding platforms, but only focused on GitHub between 2009 and 2016. Bavota \cite{Bavota2016} conducted a study about mining unstructured data in software repositories. However, this study has moved far from Consentino et al. \cite{Cosentino.2017} and Bavota \cite{Bavota2016}, by not being restricted to a coding platform or a particular technique.

By carrying out a systematic mapping study, this work upgraded \textit{the original MSR Cookbook} recommendations that are still valid and was able to make nine new recommendations on the basis of the comments in the selected papers. Moreover, it has mapped new tools that are employed in mining techniques, as well as listing the lessons learned from the current and previous findings. 

\section{Methodology} \label{sec:methodology}

This section describes the research methodology employed in this work. A systematic mapping study was first carried out 
based on the procedures defined by Petersen et al. \cite{Petersen.2015}. It describes the goals, search strategy, data extraction and analysis. Having selected the primary studies, it follows the procedural stages for building the extended Cookbook based on the \textit{the original MSR Cookbook} \cite{Hemmati.2013}. We also examine the raw data extracted, together with all the categorizations defined to encourage reproducibility \cite{xls.ZENODO}. 

\subsection{Goal} \label{sec:methodology.goal}

\begin{figure}[h]
    \centering
    \includegraphics[width=0.70\linewidth]{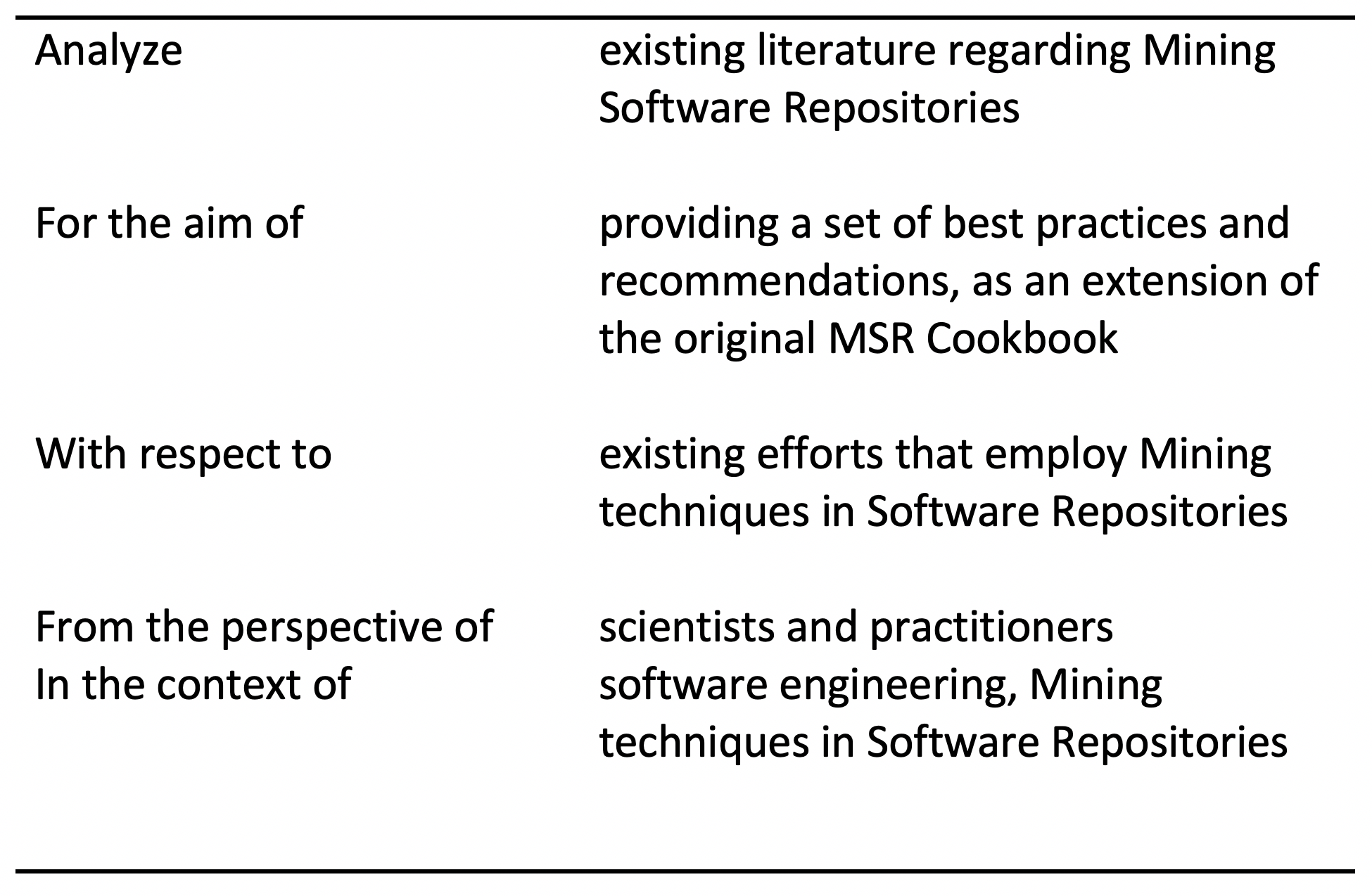}
    \caption{Goal definition for the systematic mapping study}
    \label{fig:GoalDefinition}
\end{figure}

One of the key stages in a systematic mapping study is defining the goal 
of the research which is undertaken in this study by adopting the goal definition model 
of Basili et al. \cite{Basili.1994} as shown in Figure \ref{fig:GoalDefinition}.

On the basis of the definitions shown in Figure \ref{fig:GoalDefinition}, this study raises the following research question (RQ): \textbf{What high-level themes do the primary studies address?} The RQ must divide the selected studies into the themes proposed by \textit{the original MSR Cookbook}, which are those explained in Section \ref{sec:relatedwork}. Furthermore, it would be useful to track the evolutionary pattern of current trends in research, as well as to identify any gaps and make further recommendations to provide more extensive knowledge on the topic.

\subsection{Search strategy}  \label{sec:methodology.search}
The search strategy comprises four steps: a) selecting the digital libraries of primary studies; b) defining the selection criteria; c) defining 
the search string; and d) establishing the search procedure.

\textbf{Digital libraries} were selected as long as they complied with the following the conditions: (i) number of indexed and published studies, (ii) they matched the subject with the main goals of this paper, (iii) availability of publications, and (iv) the feasibility of finding studies that made proper use of the advanced search. By conforming to the conditional requirements and the work of Dybå et al. \cite{Dyba.2007}, this study made use of the key digital libraries for Software Engineering, which were as follows : ACM Digital Library\footnote{\url{https://dl.acm.org/}}, IEEE Xplore\footnote{\url{https://ieeexplore.ieee.org/}}, and Wiley Online Library\footnote{\url{https://onlinelibrary.wiley.com/}}. Springer Link\footnote{\url{https://link.springer.com/}} and ScienceDirect\footnote{\url{https://www.sciencedirect.com/}} digital libraries, both recommended by Dybå et al., were not included as they do not allow an "Abstract" search and advanced filtering. This meant they could not qualify as they were unable to conform to the condition (iv).

The \textbf{selection criteria} were based on inclusion (IC) and exclusion constraints (EC). A study is included if: (IC1) The study is related to Software Repositories and uses Mining techniques. On the other hand, a study is excluded if: (EC1) The study was published before 2013, as in the case of \textit{the original MSR Cookbook} which covered MSRConf studies between 2004 and 2012; (EC2) The study is not written in English; (EC3) the study is a poster, tutorial, dissertation, thesis or editorial; (EC4) The study is a previous version of the same subject as that of this research; (EC5) The study is secondary or tertiary (e.g., reviews, surveys); (EC6) The study does not deal with Software Repositories; (EC7) The study is related to Software Repositories but does not apply any Mining techniques; or (EC8) the study is duplicated in one or more digital library.

A \textbf{search string} was created with the aim of mapping the MSR state-of-art and returning studies in an assertive manner to categorize them accordingly. During the pilot study, we discovered that recent MSR papers are likely to use Artificial Intelligence or Machine Learning as devices to explain the mining process in the papers. 
The search string included the terms "Mining", "Learning", and "Artificial Intelligence" 
with "Software Repositories" and their synonyms. Owing to the fact that one of the most important conferences (MSRConf) has common names in the search string, the abstract was the only field included in the search within the databases chosen. As in the case of the pilot study, the metadata that were found ignored the search constraints and took all of the MSRConf papers into account when only filtering by "title" or "keywords". The search string chosen for this study was: \textit{("Mining" AND "Software Repositories")} OR \textit{("Mining" AND "Software Repository")} OR \textit{("Mining" AND "Public Repositories")} OR \textit{("Mining" AND "Public Repository")} OR \textit{("Learning" AND "Software Repositories")} OR \textit{("Learning" AND "Software Repository")} OR \textit{("Learning" AND "Public Repositories")} OR \textit{("Learning" AND "Public Repository")} OR \textit{("Artificial Intelligence" AND "Software Repositories")} OR \textit{("Artificial Intelligence" AND "Software Repository")} OR \textit{("Artificial Intelligence" AND "Public Repositories")} OR \textit{("Artificial Intelligence" AND "Public Repository")}.

A \textbf{search procedure} was established to select the primary studies and then answer the research question. This procedure followed three stages: (i) selecting the studies in the digital libraries chosen; (ii) applying the selection criteria, subject to the inclusion and exclusion constraints; and, (iii) extracting and analyzing relevant information from the selected studies. The search procedure is divided into the following phases:  Phase 1 applies the search string to the digital libraries chosen, following the selection criteria without any exclusion. Phase 2 filters the raw dataset by means of the basic qualitative exclusion criteria EC1/2/3/4/5 to get the first primary dataset. Finally, Phase 3 applies the exclusion criteria EC6/7/8, (which are regarded as the most accurate), and the results obtained from the final dataset of the selected studies.

\subsection{Data Extraction} \label{sec:methodology.extraction}

The data extraction process collects each item of Extracted Information (EI) from the digital libraries as follows: (EI1) the affiliation of the authors, given by the author´s name, university and country; (EI2) the main details of the study , including the title, publishing date, and DOI [Digital Object Identifier]; (EI3) Type of material, whether the study is an article, a research article or a tool; (EI4) Applied Mining Technique, whether the applied technique is a Supervised Mining Technique, Unsupervised Mining Technique, Evaluative Analysis of Mining, Machine Learning, Deep Learning, or any other technique or heuristic chosen; (EI5) the theme addressed by the study and categorized by this study.

\subsection{Data Analysis} \label{sec:methodology.analysis}

After the search procedure, there was an in-depth examination of the selected papers to derive the recommendations per theme. Here, we followed \textit{the original MSR Cookbook} by interpreting a recommendation of a theme as representing a consolidation of key features  (called comments) found in the list of papers. This involved  first carrying out open coding and axial coding procedures~\cite{Strauss.Corbin_1998} to qualitatively analyze the topic that resulted in comments, which were later reviewed and incorporated  in  the recommendations. Owing to the number of extracted comments, we again followed \textit{the original MSR Cookbook} and decided that relevant recommendations would be those that were found in at least two selected papers. Two researchers worked together, where each one worked on a subset of the recommendations extracted from the selected studies. The researchers discussed all classifications until they reached a consensus about the results. Another two experienced researchers got the subsets processed to review them and refine the mapping between the theme and recommendations. In addition, we held a weekly meeting with the whole group of co-authors to discuss and adjust results until we reached an agreement. This study did not calculate the Inter-rater reliability (IRR) as the purpose of mapping recommendations was to find categories (themes) rather than classify the data according to emerged codes \cite{irrPaper}.

\section{Results} \label{sec:results}

This section first characterizes the results of this study, and later introduces the recommendations of the themes. 

\subsection{Studies characterization} \label{sec:results.studiescharacterization}

\begin{figure}[h]
    \centering
    \includegraphics[width=0.38\textwidth]{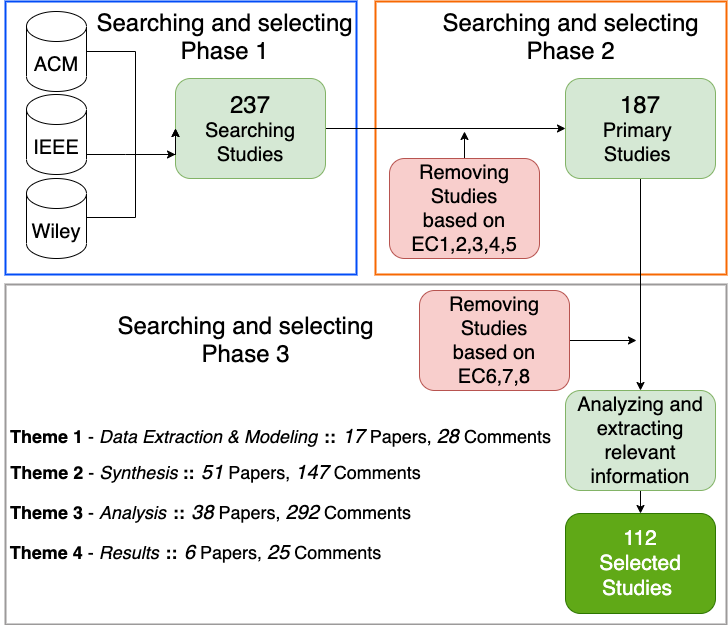}
    \caption{Searching and selecting procedure of studies}
    \label{fig:SelectedStudies}
\end{figure}

The data extraction was carried out between September and November 2020. With regard to the procedures involved in the search and selection, in Phase 1 the search returned 237 studies - the majority obtained by IEEE, followed by ACM and Wiley. Phase 2 applied EC1/2/3/4/5 exclusion criteria with an outcome of 187 studies, and the percentage of returned studies remained about  the same, i.e., IEEE returned most of the studies, followed by ACM and Wiley. Lastly, the remaining exclusion criteria (i.e., 6, 7, and 8) were applied in Phase 3 with a final list of 112 selected studies and, although there was less  difference between IEEE and ACM, IEEE was the data library with the most studies, followed by ACM and Wiley. Figure \ref{fig:SelectedStudies} outlines the search and selection procedure 
per Phase and shows the number of selected papers and comments per the theme. 

Apart from the details of the papers, obtained from the 112 selected studies, the data extraction process enabled a considerable number of comments to be created for each theme (492 in total) that later culminated in recommendations. 

\begin{figure}[h]
    \centering
    \includegraphics[width=0.70\linewidth]{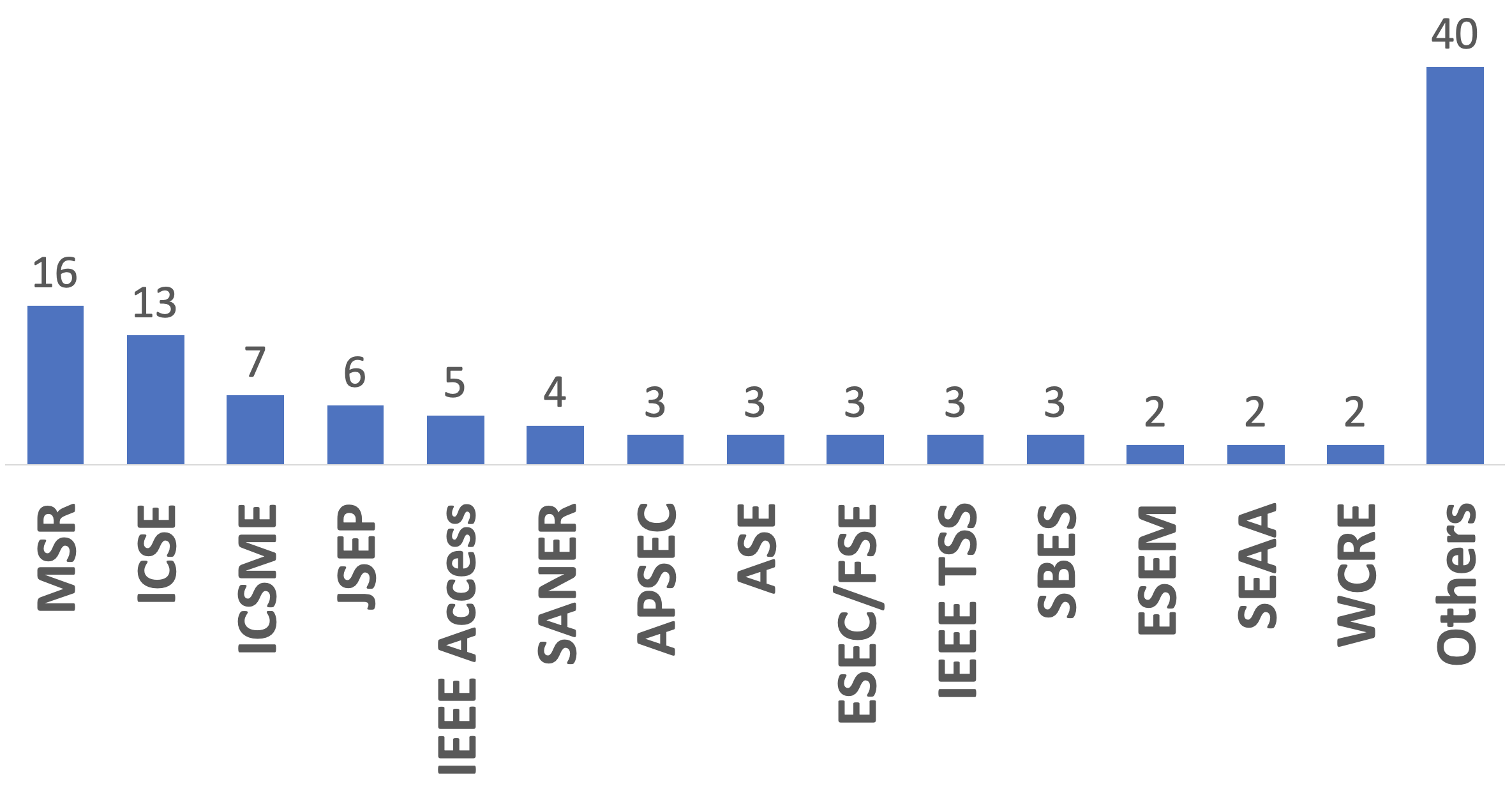}
    \caption{Selected studies from each venue}
    \label{fig:StudiesVenues}
\end{figure}

With regard to the venues of the selected papers, the MSRConf was the conference which had most published papers, followed by The International Conference on Software Engineering (ICSE)\footnote{\url{http://www.icse-conferences.org/}}, The International Conference on Software Maintenance and Evolution (ICSME)\footnote{\url{https://conferences.computer.org/icsm/}}, and The Journal of Software: Evolution and Process (JSEP) \footnote{\url{https://onlinelibrary.wiley.com/journal/20477481}}. This demonstrates that our mapping study was able to identify papers from significant publication venues in the area of software engineering. Figure \ref{fig:StudiesVenues} shows a graphical representation of the distribution of papers among the publication venues.

\subsection{Themes Recommendations} \label{sec:results.themesrecommendations}

This section outlines the updated and new recommendations for each high-level theme of the extended MSR Cookbook. The number of comments that were found in the selected papers of this study is also shown as a means of demonstrating the significance of the recommendation itself, as well as how it fits the theme.

In the next sections, the following scheme is adopted to 
introduce the recommendations and identify each of the themes. The first letter denotes its theme -- "D" for Data Extraction and Modelling (Theme 1), "S" for Synthesis (Theme 2), "A" for Analysis (Theme 3), and "R" for Results (Theme 4). While the second letter indicates whether the recommendation belongs to \textit{the original MSR Cookbook} by using the letter "O" (Old) or is something new - through the letter "N" (New). These two letters are combined with a sequential number which is used to order the recommendation within a theme (i.e., DO1 should be understood as the first recommendation of Theme 1 and an updated version of \textit{the original MSR Cookbook}). The recommendations shaded with a gray background and a solid line in a frame box were those found in \textit{the original MSR Cookbook}, while the ones shaded in green 
with a dashed line represent the new additions made to this study. As this study is an extension, \textit{the original MSR Cookbook} recommendations that did not appear in this study will be only shown in the Figure \ref{fig:MRS.Past.Present} as past references.

\subsubsection{Theme 1: Data Extraction and Modeling} \label{sec:results.themesrecommendations.theme1}

This theme was an important topic in past research studies \cite{Hemmati.2013}. However, recent study trends have been more concerned with applying techniques to get relevant information, rather than extracting and preparing data for usage. This can be explained by the technological maturity of the MSR research tools available for extracting and preparing data for  the next themes (2-4) \cite{Meurice.2014}. In other words, our findings suggest that there are already mature tools that are good enough  to perform data extraction and modeling, and hence there has been a decline in number of published papers on Theme 1 for some time.

Source Control Management (SCM) is one of the key features of MSR. The domain knowledge of SCM is essential for extraction and preparing data mining tools, as well as the heuristics for scheduling data gathering \cite{German.2015}, and enhancing existing process capabilities \cite{Gupta.2014}. This is established in the following recommendation. 

\hspace*{-0.4cm}\includegraphics[width=8.5cm]{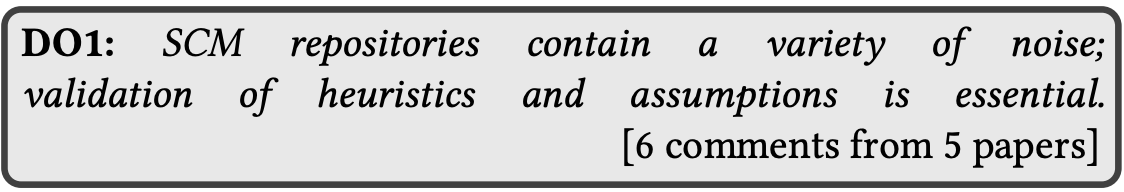}

Source code mining can be either a straightforward or costly process since it requires  knowledge and time to extract relevant information, which thus influences the choice of code extraction method. Regular expressions for mining data in large software repositories can be, for instance, an alternative to reducing  data collection significantly \cite{Bakar.2014}. In contrast, tools and abstractions can hide the source of complexity by turning a complex process of code mining into a simple query mechanism \cite{Tiwari.2016}. 

\hspace*{-0.4cm}\includegraphics[width=8.5cm]{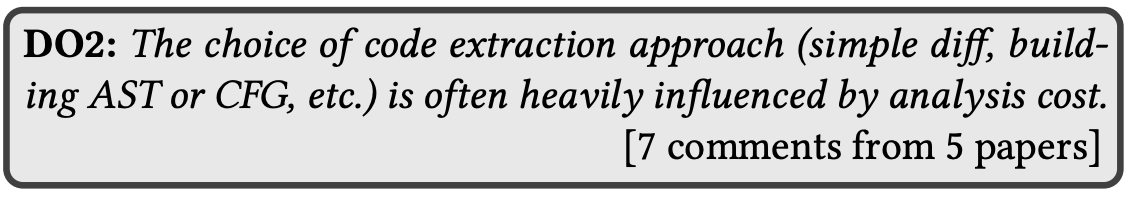}

The link between software repositories and external data sources (Heterogeneous Data Systems (HDS)) represents a complex feature of MSR \cite{Wilder.2016}. Domain-Specific Languages (DSLs) can act as an interface to access and connect HDS into the software repository mining \cite{Wilder.2016}, by reducing the complexity of MSR and providing new insights. However, owing  to domain complexity, the construction of DSL can be a hard process. Some studies have addressed 
this issue by conducting a Feature-Oriented Domain Analysis to overcome the challenges more easily  \cite{Huang.2013}. Thus, this work makes the following remarkable recommendation in Theme 1. 

\hspace*{-0.4cm}\includegraphics[width=8.5cm]{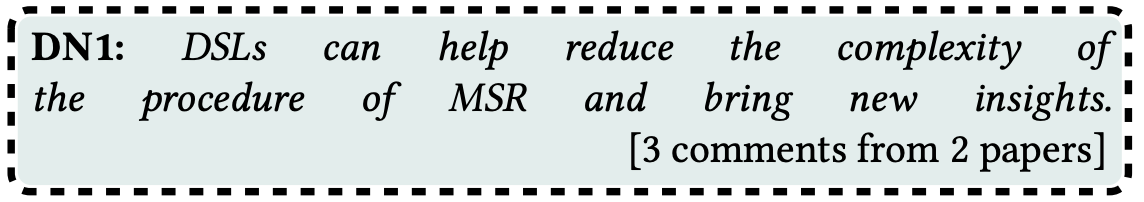}

There is a common sense view that the history of database systems helps our understanding of mining approaches, including MSR. Software development has become increasingly complex, by aggregating different sources of information such as defect tracking systems. Hence, some studies have suggested  ways to incorporate additional information into the MSR process, such as adding issue tracking for databases \cite{Steinbeck.2020} or even providing a dataset for bug mining information \cite{Lamkanfi.2013}. As the way this database is evolving has the potential to contribute to MSR, this work also makes a  supplementary new recommendation on Theme 1.  

\hspace*{-0.4cm}\includegraphics[width=8.5cm]{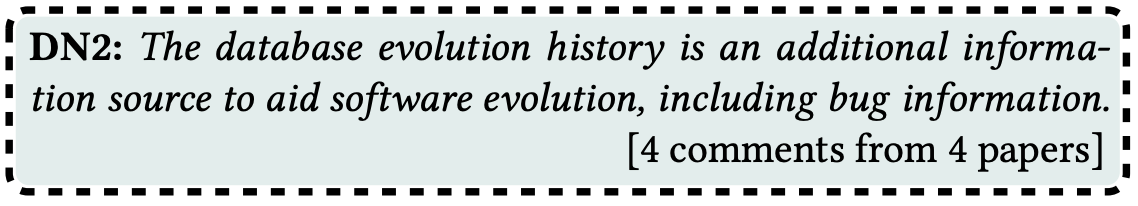}

\subsubsection{Theme 2: Synthesis} \label{sec:results.themesrecommendations.theme2}

This theme focuses on the application of mining techniques, such as clustering and regression, and papers about it have been continuously published in recent years. Our study does not show a significant reduction in publications in Theme 2. Likewise, there remains a great interest in the practice and research, which is encouraging new studies 
to be undertaken.

One way that is recommended to improve the chances of obtaining a decent result in a mining algorithm, is to preprocess data properly before the mining task. This is why Theme 1 comes before Theme 2. Hence, one useful recommendation in this theme is to tune parameters and perform a sensitivity analysis, as they can improve the modeling process. It is also possible to find works doing tasks such as comparing text-based and dependence-based approaches for determining the origins of bugs \cite{Davies.2014}. 

\hspace*{-0.4cm}\includegraphics[width=8.5cm]{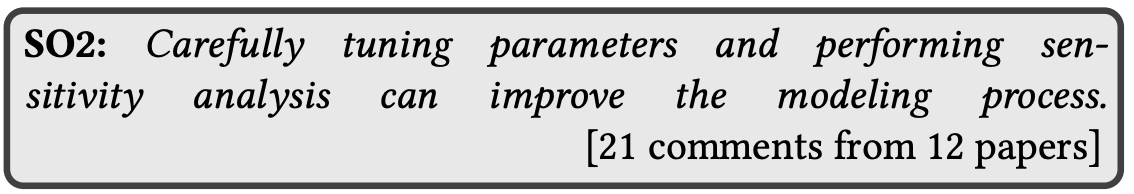}

In MSR, a combination of classifiers (ensemble of classifiers) is likely to exceed the accuracy of a single classifier by combining several of them. There are works in the literature that adopt this approach and provide a taxonomy for ensemble-based methods to overcome problems of an imbalanced classification \cite{Galar.2012}. The combination of results can be visualized in some MSR tools \cite{Barros.2020}, where the main indicators (such as accuracy and prediction) can be compared, and thus allow better decision-making. This is a key factor and our study makes a new recommendation on this theme.

\hspace*{-0.4cm}\includegraphics[width=8.5cm]{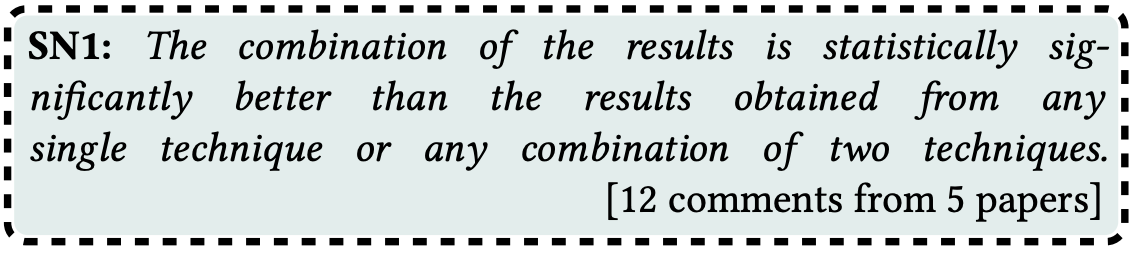}

Artificial Intelligence (AI) is definitely a trend in MSR papers \cite{Luzgin2020} and the performance standard  of its solutions is related to how good the data source is. Since it can be expected that there will be plenty of data to work with, rich domain knowledge is essential for AI solutions in terms of their performance and evolution. A common comment in papers that adopt this approach is the relationship between rich domain knowledge and success, such as data-driven applications \cite{Misra.2017} or a framework for the retrieval of information \cite{Parizy.2014}. Strategies to keep software projects accurate and cohesive, such as improving software maintenance \cite{Gupta.2017} or defect prediction \cite{Abdalkareem.2020}, are linked to rich domain knowledge as well. 

\hspace*{-0.4cm}\includegraphics[width=8.5cm]{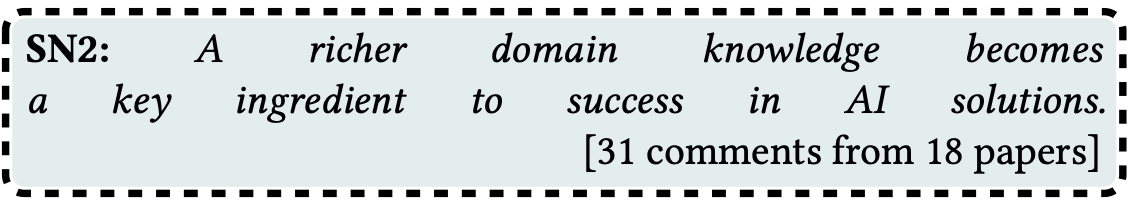}

The purpose of  feature selection algorithms is to obtain the most suitable subset from the source that fits the prediction model best, by removing redundant and irrelevant data, to ensure  the best possible prediction \cite{Guyon.2003}. Feature selection outcomes show impressive figures for prediction such as entropy and gain, and there are cases where the merging of  feature selection and other techniques proved to be successful, such as learning-based approaches \cite{Yue.2018} or the longest common sequences (LCS)\cite{Neysiani.2019}. However, applying feature selection filtering to just the highly ranked features does not necessarily improve  the performance of the measures \cite{Abozeed2020}.

\hspace*{-0.4cm}\includegraphics[width=8.5cm]{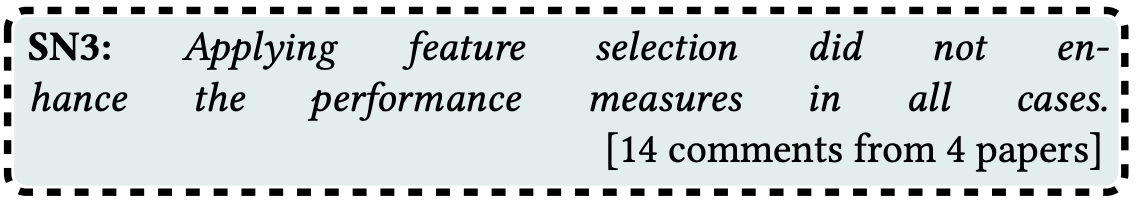}

Despite the fact that when combined in a clever manner, feature selection  can definitely enhance the prediction models, when compared with recently extracted features,  historically-based features can make a significant improvement. Learning-based \cite{Yue.2018, Abdalkareem.2020} and ranking \cite{Ye.2019} approaches are outperforming previous models that  were not based on historical data.

\hspace*{-0.4cm}\includegraphics[width=8.5cm]{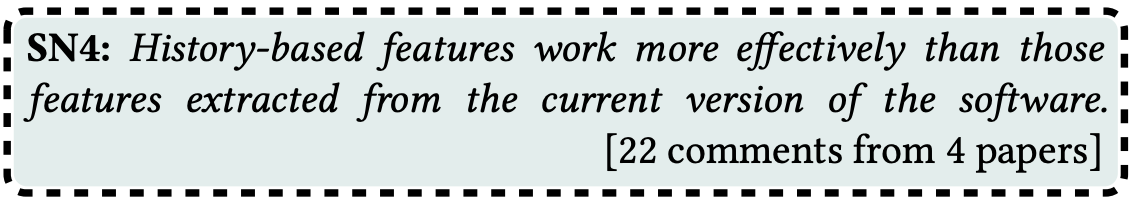}

Deep Learning has achieved astonishing results in many areas of research, including MSR, and progress is underway \cite{Abozeed2020}. In fact, studies show that Deep Learning has been obtaining a higher degree of accuracy than other techniques
. For instance, the automatic categorization of software projects through MSR \cite{AnhTuanNguyen2017}, or even bug prediction models \cite{Abozeed2020} are relying on Deep Learning as the main tool to obtain good  results.

\hspace*{-0.4cm}\includegraphics[width=8.5cm]{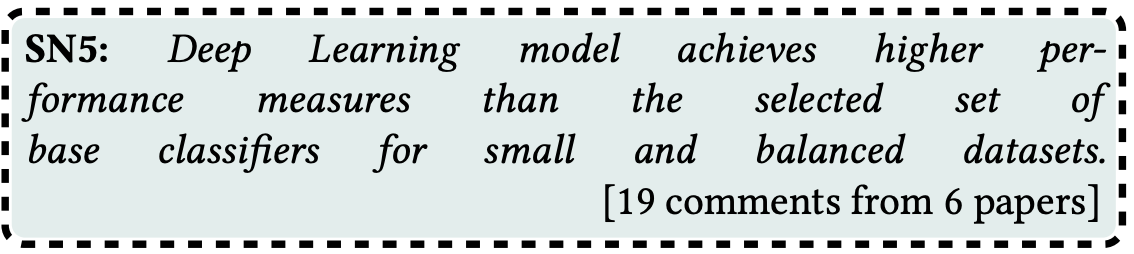}

\subsubsection{Theme 3: Analysis} \label{sec:results.themesrecommendations.theme3}

In MSR, the analysis of results is definitely one of the most important phases, since the conclusions are the main output. For this reason, Theme 3 is still one of the subjects which has attracted the most attention from  researchers, and been  the source of most of the  comments found in this study.

Mining algorithms and processes have relied on automated tasks, where the means of getting the best result can show wrong figures or biased results. In order to address this challenge, manual verification of all outputs should be presented in an analysis of MSR approaches. For instance, a manual analysis of the relationship between refactoring and bugs should be done, as refactoring always encourage bug-fixing activities \cite{Penta.2020}. However, the task of manually checking all of the  outputs can be very demanding and become impossible to achieve. Some automated sampling and verification strategies are suggested
, such as a machine learning-based model which includes efficient preprocessing together with highly accurate prediction and a low level of errors \cite{Tariq.2020}.

\hspace*{-0.4cm}\includegraphics[width=8.5cm]{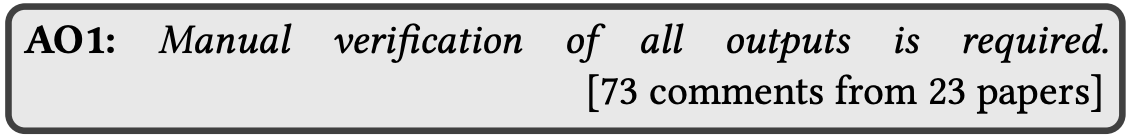}

The evolution of data mining techniques can be seen in recent works, and this has led this research to conduct a new systematic mapping study as one of its goals. Before applying sophisticated mining approaches, some studies have found simple and efficient ways to check hypotheses about a set of data. Others can answer simple questions through a  correlation analysis before taking a further step into mining techniques, such as determining  whether bugs foreshadow vulnerabilities \cite{Camilo.2015}.

\hspace*{-0.4cm}\includegraphics[width=8.5cm]{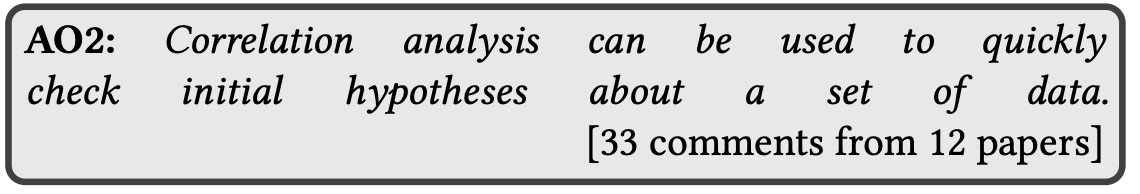}

Precision and recall as the main measures of mining approaches have been adopted by researchers and practitioners with a strong label of trust. For this reason, the majority of papers include precision and recall measures to ensure success in their strategies and experiments. Nonetheless, in certain contexts, precision and recall may not be enough, and some studies use F1-Score, Recall \cite{Barros.2020}, or even techniques such as PCA and Dynamic Weighting \cite{Ali.2013}.

\hspace*{-0.4cm}\includegraphics[width=8.5cm]{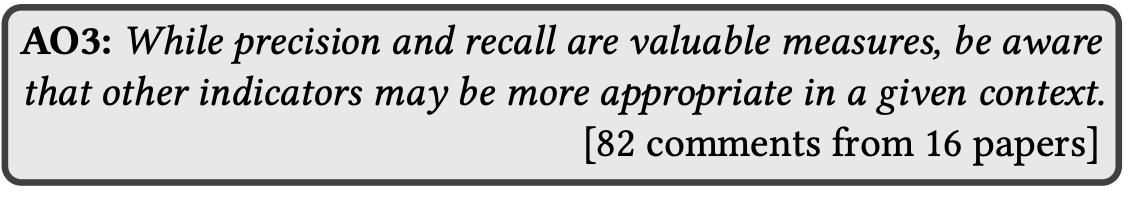}

In a mining approach, statistical analyses are extremely important to understand state-of-the-art metrics and to determine why some metrics are relied on more than others. If conclusions are only drawn on the basis of statistical analyses, this can blur a proper understanding of the data mining system. Practical differences, based on empirical evidence such as the impact of refactoring on the relationship between quality attributes and design metrics, can lead to more assertive conclusions rather than simply providing a complex statistical analysis \cite{AlOmar.2019}.

\hspace*{-0.4cm}\includegraphics[width=8.5cm]{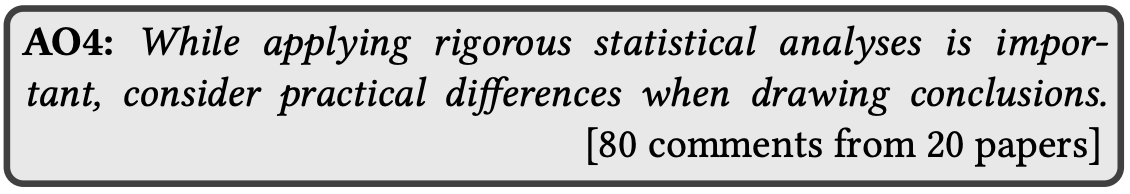}

Managing resources in software projects is one of the most challenging tasks. Planning, scheduling, and allocating resources can be very tricky, especially if the estimations are only based on random heuristics or guesses. However, the prediction of software effort can really help in managing resources in software projects, which has been demonstrated in studies such as those that entail selecting the right predictors from huge software repositories in order to ensure an accurate software effort estimation \cite{Tariq.2020}.

\hspace*{-0.4cm}\includegraphics[width=8.5cm]{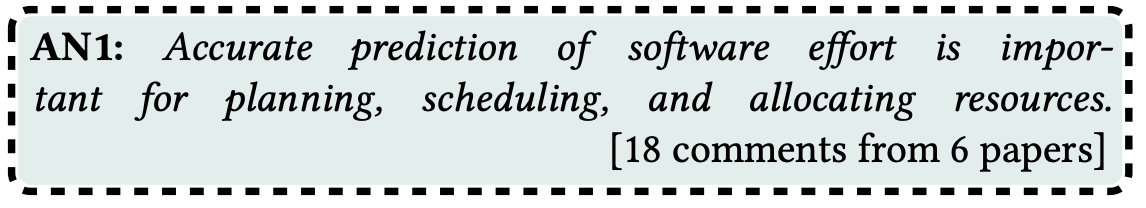}

\subsubsection{Theme 4: Results}
\label{sec:results.themesrecommendations.theme4}

This theme corresponds to results where sharing data and outcomes permit a further validation and replication of the studies. MSR studies generally do not possess enough evidence to allow their individual results to be applied in other studies, as the context or data sources might be very different. For this reason, this theme, which is often ignored by most researchers, is very important in the MSR field to keep the scheduling and replication of these studies alive. Oddly enough, Theme 4 still has very few publications and although there is an increase of Theme 4 papers in this study, they are far from reaching their full potential.

The feasibility of Theme 4 depends on sharing studies data, tools, and techniques to allow reproducibility. Historically, the replication of studies in MSR has not been straightforward. 
Here, in recent years there have been a number of publications that have enhanced the datasets with cleaning and preprocessing text data \cite{Lazar.2014}, along with frameworks for a replication analysis of software projects \cite{Ghezzi.2013}.

\hspace*{-0.4cm}\includegraphics[width=8.5cm]{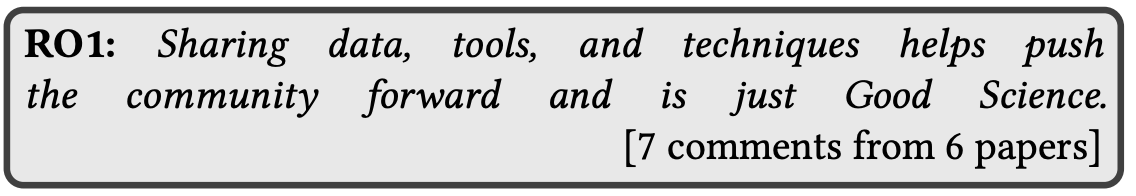}

As long as replication studies are available, it should be a straightforward task to compare mining techniques or assess findings in many projects. An example of replication is the datasets available from studies, where bugs extracted from tracking systems are mined and compared with the sources of different systems \cite{Lazar.2014}.

\hspace*{-0.4cm}\includegraphics[width=8.5cm]{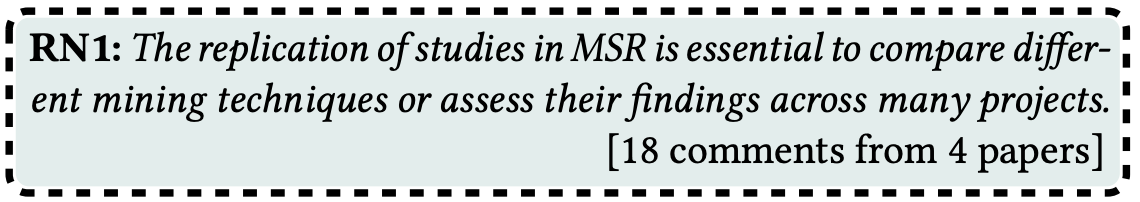}

\section{Discussions} \label{sec:discussions}


\subsection{Implications for literature and practice} \label{sec:discussions.implications}

The research contributions made by this work include the following: 1) an update of the MSR field, carried out by a systematic mapping study; 2) a wider coverage of MSR studies; and 3) an extended Cookbook for updating recommendations, making tools available, and providing lessons learned for future research. Briefly, \textbf{an update of MSR research is needed as it is an area that is  constantly growing}. To address this, this work reviews the existing state-of-the-art and draws attention to the challenges facing researchers and practitioners. \textbf{As the spread of MSR research has now become a reality, a wider coverage of studies is absolutely indispensable}. Here, rather than restricting our review to the MSRConf as the works in the literature have done, this study took into account all of the  digital libraries available for mapping, including several conferences, workshops, and new works in the MSR field. Figure \ref{fig:StudiesVenues} summarizes the main publishing venues, where major international conferences obtained most of their selected studies, which underlines the importance and heterogeneity of this study. \textbf{Owing  to the heterogeneity of MSR studies, grouping these studies is essential.} This work extended and upgraded the \textit{the original MSR Cookbook} to divide the  studies into four high-level themes. This is designed to ensure that new research is conducted properly and keeps the focus on MSR as a development field. The measures taken are discussed in further detail in the next sections. 

\subsection{RQ analysis: In what high-level theme are the primary studies addressed?} \label{sec:discussions.rqanalysis}

Theme 2 obtained the majority of papers, followed by Theme 3, Theme 1, and finally Theme 4. In fact, the systematic mapping study revealed that the papers on Theme 1 are declining  and may reach an almost zero output over time, which suggests that data extraction and modeling is a mature subject in the MSR field. Another sign of MSR maturity is the increase in the number of Theme 4 papers, which are still under-represented relative to the others, but has increased when compared with \textit{the original MSR Cookbook} \cite{Hemmati.2013}. 

\begin{figure}[h]
    \centering
    \includegraphics[width=0.70\linewidth]{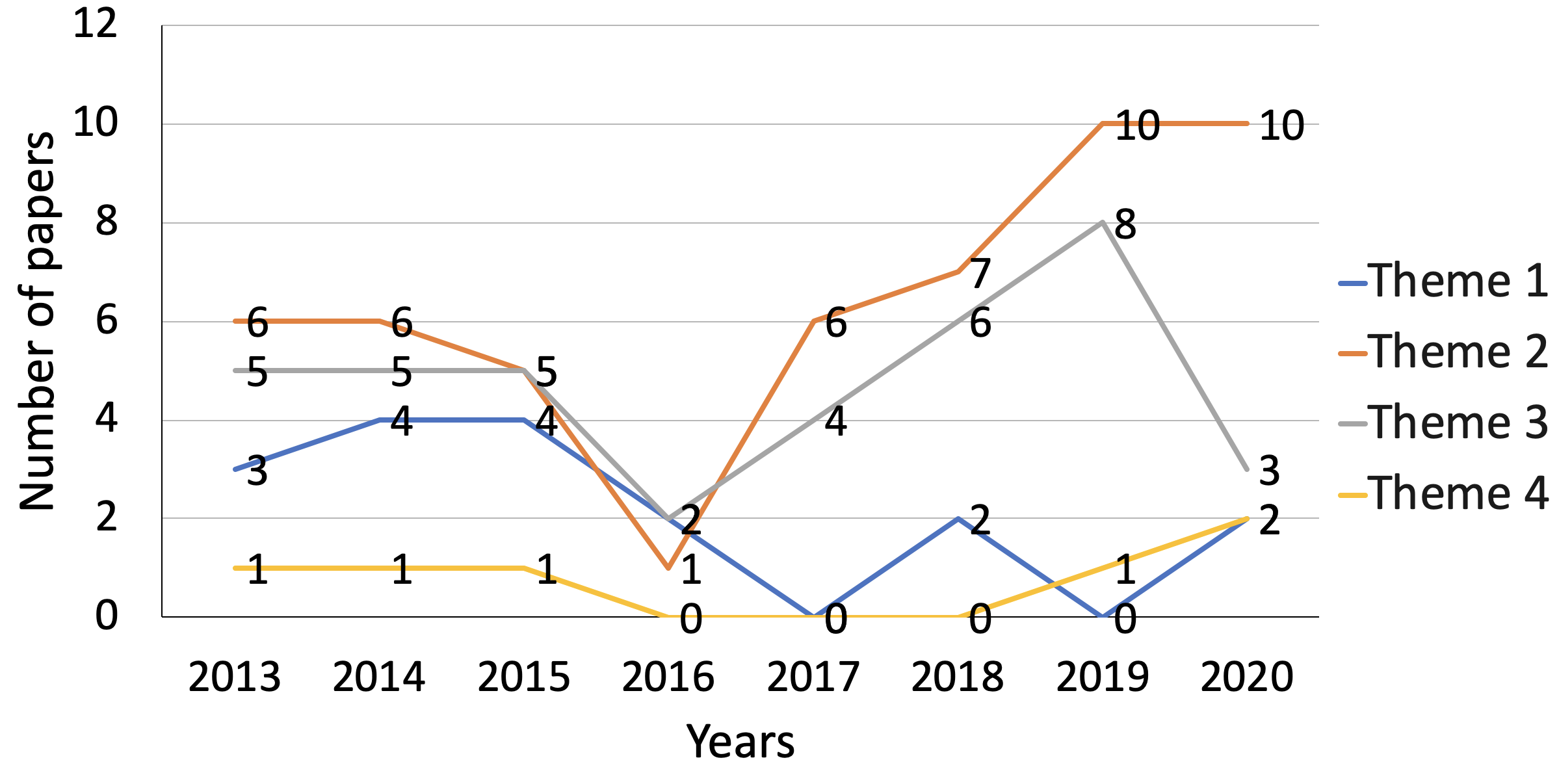}
    \caption{Evolution of selected studies themes}
    \label{fig:EvolutionOfThemes}
\end{figure}

Figure \ref{fig:EvolutionOfThemes} plots the evolving pattern of selected studies on the themes over time. Even though a few papers were found, the number of studies in Theme 1 is decreasing. In contrast, Theme 2 has been growing in recent years, especially since 2016, which suggests that Theme 2 has become a trend in the MSR field since then. This might be due to the fact that data mining applications have space in the literature and, as the MSR field evolves, AI is becoming  a reality on MSR and might be the next growing subject in the years ahead. On the opposite side, research studies with regard to Theme 3 have decreased in the last few years, which suggests  that an analysis of implemented techniques is no longer common. Last but not least, the number of publications on Theme 4 is increasing.

\subsubsection{Evolution of recommendations} \label{sec:discussions.evolutionofrecommendations}

\begin{figure}[h]
    \centering
    \includegraphics[width=0.45\textwidth]{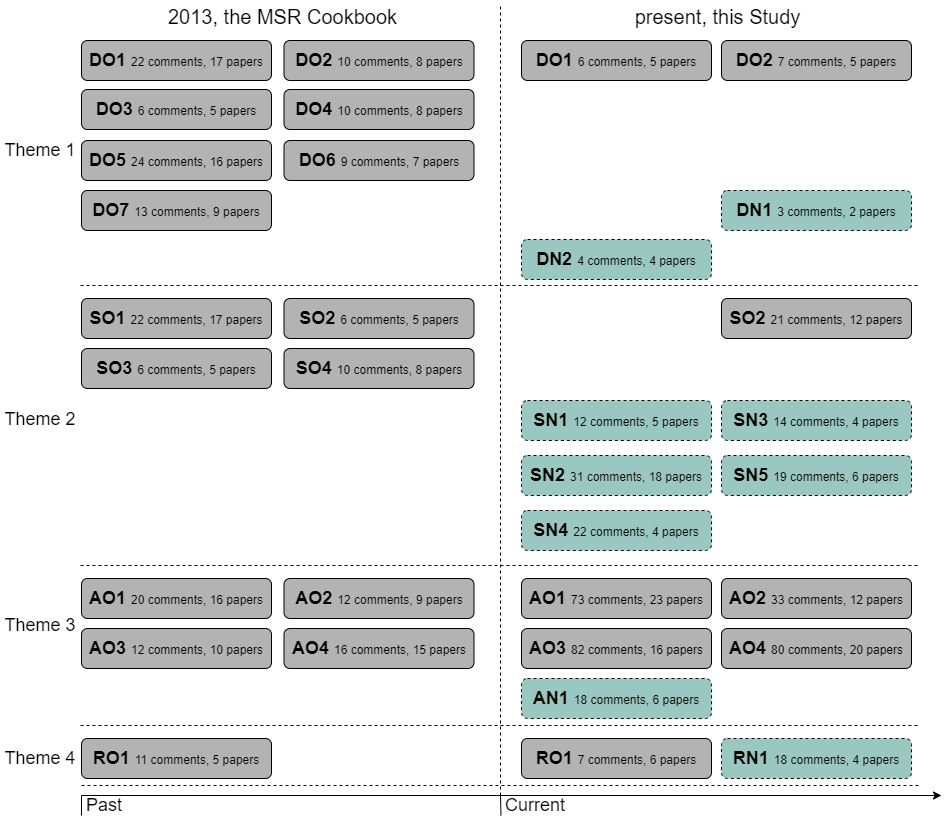}
    \caption{Evolution of themes recommendations}
    \label{fig:MRS.Past.Present}
\end{figure}

With regard to the evolving pattern of  recommendations, when  this study is compared with \textit{the original MSR Cookbook}, Figure \ref{fig:MRS.Past.Present} shows how this has taken place. The relative position of the original recommendations is kept in the diagram to allow a graphical comparison to be made, even if they were not found in either previous studies or in this work. When our findings were compared with \textit{the original MSR Cookbook}, there was a significant decrease in recommendations in Theme 1 that can  be explained by the fall in the number of  papers in recent years. Even though there remained previous recommendations about SCM and code extraction, two new recommendations appeared which showed how MSR  had evolved. The first (DN1) suggested ways of  how to perform MSR more easily, while the second (DN2) focused on how historical data  can lead to the evolution. Theme 2 kept the recommendation of \textit{the original MSR Cookbook} in the area of tuning parameters and conducting a sensitivity analysis (SO2), and there was a significant increase in this recommendation. The new recommendations this study found in Theme 2 showed the combination of techniques required  to achieve better results, as well as those related to AI techniques, i.e., a richer domain knowledge; feature selection; historically-based features; and deep learning models. Theme 3 was given a new recommendation, related to the prediction of software effort, and kept the previous ones. Finally, the single previous recommendation of Theme 4 was kept as well, 
with an increase in the number of its comments, although a new recommendation about the replication of studies was also identified. 

\subsection{The Tools of MSR} \label{sec:discussions.thetoolsofmsr}

MSR is also driven by the creation of tools to assist in carrying out tasks and improving correlated  processes. Spadini et al. \cite{Spadini.2018} created the tool PyDriller, a Python framework that makes it easier to extract data from mining Git repositories. The Boa \cite{Dyer.2013} tool provides snapshots of Java projects in GitHub and SourceForge and is designed to reduce the complexity of mining. Other tools are concerned with helping to integrate software development with other areas, such as the gthbmining \cite{Barros.2020} tool where the main goal is to discover DevOps trends in public repositories. There are also papers focused on giving advice on  how MSR should be undertaken. Kalliamvakou et al. \cite{Kalliamvakou.2014} performed an empirical study about public repositories and how contributors should interact with them.

The tools obtained from the selected studies are categorized as well, to strengthen the applicability of the extended Cookbook. Although categorizing tools can be very difficult, the main purpose of this tool mapping is to help the MSR community to make the best choice to stimulate new experiments and theories, by allowing researchers more time to devote to making  advances in science, instead of creating new tools that have been designed  and published before. Table \ref{fig:MSRTools} displays the tools collected in the selected studies grouped by themes and suggested categories, as well as the number of studies that were found. Further details can be found at the replication package \cite{xls.ZENODO} (sheet "Tools").

\begin{table}[h]
    \centering
    \caption{Summary of Tools used in each Theme}    \includegraphics[width=0.705\linewidth]{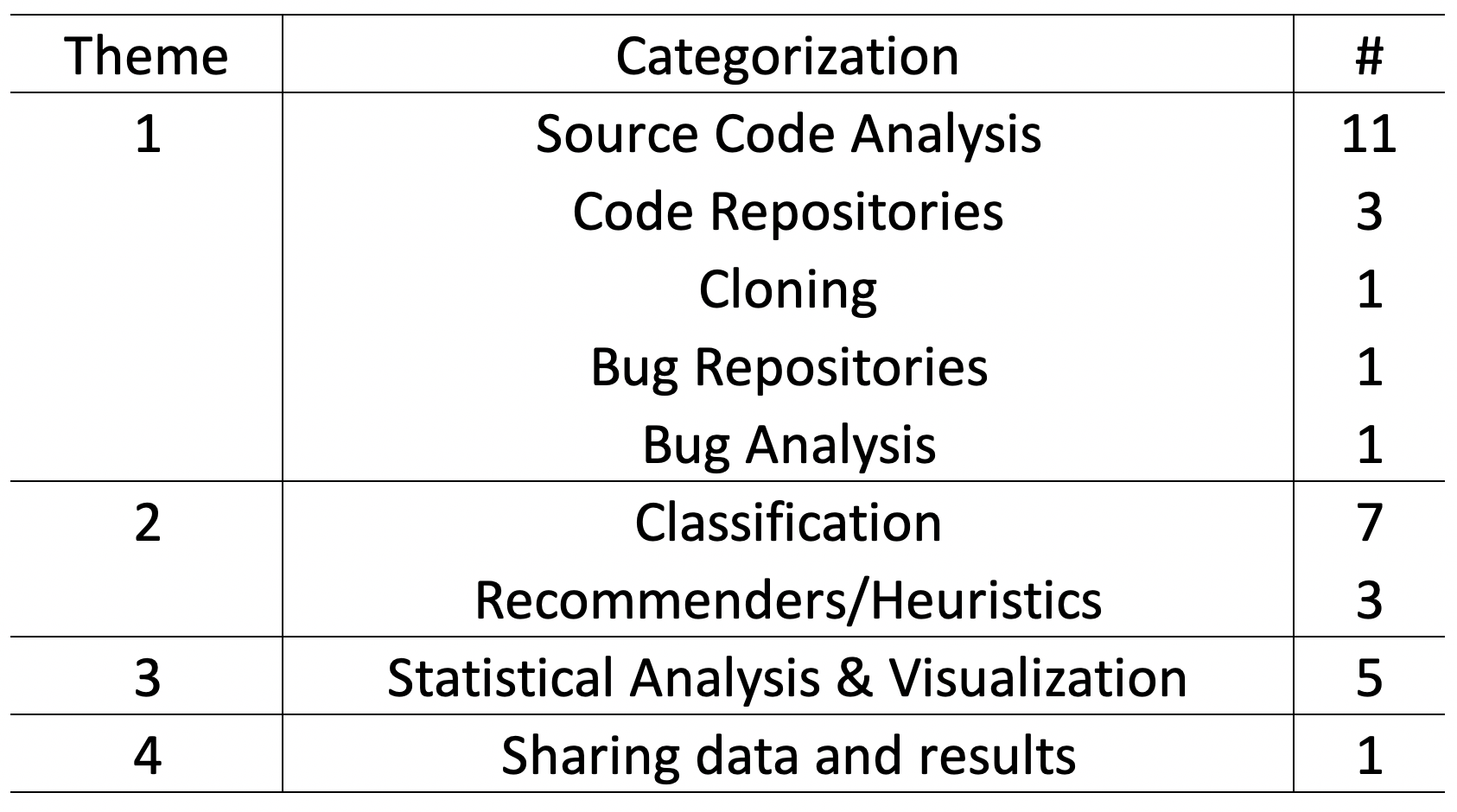}
    \label{fig:MSRTools}
\end{table}

\subsection{Lessons learned} \label{sec:discussions.lessonslearned}

The results and mapping of the evolution of recommendations allowed this study to state what lessons were learned from the existing literature. As our findings showed that the themes and upgraded recommendations still coincided with the four high-level themes of a typical MSR process, the lessons learned were also grouped by themes. Finally, the findings  
from related works are provided, as a means of  guiding future extensions of this Cookbook.

Theme 1: \textbf{Noise is still a challenge in data extraction and modeling.} Noises can be found not only in the source code (as pointed out by Recommendation DO1 \cite{Guilardi.2020}), but all kinds of software repositories (i.e. version control systems, bug trackers, and communication channels). Thus, one of the first steps of data extraction should be cleaning up data to prevent noises by using heuristics and taking into account the results of a cost analysis (Recommendation DO2 \cite{AwangAbuBakar.2014}); \textbf{The complexity of the procedure of MSR can be reduced.} The MSR process of data extraction and modeling can be made easier by using APIs and tools or even drawing on the ideas available in the literature and based on practice. There are tools based on  this mapping study, as well as ideas and shortcuts for addressing complexity properly (Recommendation DN1 \cite{Wilder.2016}); \textbf{The evolution of  database technology can help provide a software enhancement}; and, external data sources such as bug trackers and mail lists should be included in the data extraction and modeling processes (recommendation DN2 \cite{Lamkanfi.2013}).

Theme 2: \textbf{If you take care of the modeling process, you can achieve improved performance measures}. Recommendation SO2 suggested measures like tuning parameters and performing sensitivity analyses to improve the modeling process \cite{Ramsauer.2019}. Unsurprisingly, Recommendation SN5 matched high-performance deep learning models with efficient modeling processes \cite{White.2015}, and achieved  better  performance measures; \textbf{A richer domain knowledge is a key ingredient for success in MSR}. Recommendation SN2 found several samples where domain knowledge is the equivalent of  MSR success \cite{Misra.2017, Parizy.2014}. Moreover, domain knowledge helps feature selection in several ways, e.g., the SN3 Recommendation demonstrated that only relying on feature selection is not enough in all the  cases that require domain knowledge \cite{Abozeed2020}, and the SN4 Recommendation showed that historical and well-known features outperform new data \cite{Yue.2018}; \textbf{A combination of results should prevail when employing  single techniques}. Recommendation SN1 showed that a combination of classifiers (e.g., ensembles) is more assertive than a single technique or any combination of two techniques \cite{Bogdan.2014}.

Theme 3: \textbf{Before reaching  any conclusions, manual checking and an understanding of the practical differences between all the outputs are encouraged}. Hence, Recommendations AO1 \cite{Dyer.2013, Penta.2020} and AO4 \cite{AlOmar.2019} are still valid, and there was an increase of comments in both 
papers; \textbf{The context must be taken into account before selecting the appropriate analysis indicators}. Although the AO2 Recommendation showed that a correlation analysis is still used \cite{Camilo.2015}, the increase in the number of AO3 Recommendations confirmed that context should be the driving-force for selecting the appropriate analysis indicators \cite{Ali.2013}. Different classification 
strategies should be stimulated like Matthews Correlation Coefficient (MCC) \cite{Tantithamthavorn.2019} or more tests employed (e.g., the Scott-Knott test) to evaluate the statistical ranks of classifiers \cite{Ghotra.2017}; \textbf{Resources management can be supported by accurate software predictions}. As suggested by Recommendation AN1 \cite{Tariq.2020}, resource planning, scheduling, and allocation can be supported by accurate software predictions. 

Theme 4: \textbf{Sharing data and replicated studies can help make advances in science}. Sharing data, tools, and techniques definitely helps science make progress as the RO1  Recommendation  confirmed \cite{Ghezzi.2013}. Additionally, a replication of studies has helped science as well, as the RN1 recommendation demonstrated in its comparisons and assessment of advances in data mining techniques \cite{Lazar.2014}.

Related works: \textbf{MSR mapping studies should include all the relevant venues}. Although previous MSR mapping studies made a significant contribution to the literature, they only covered MSRConf papers \cite{Hemmati.2013, Farias.2016, Chaturvedi2013}. In contrast this paper showed that other relevant venues (i.e. ICSE, ICSME) have been publishing important papers and, thus, should be included in future MSR mappings; \textbf{Include as many MSR techniques as you can while mapping}. Instead of being studies based on a single coding platform \cite{Cosentino.2017} or mapping studies with a few techniques \cite{Bavota2016}, a good MSR mapping should include as many techniques as possible so that it can provide recommendations and guidelines to advance science.

\section{Threats to validity} \label{sec:threatsvalidity}


\textbf{Construction validity.} This work attempted to select the best digital libraries available for collecting research data, following the recommendations made by Dybå et al. \cite{Dyba.2007}. However, owing to the constraints imposed by the search criteria, the digital libraries \textit{Springer Link} and \textit{ScienceDirect} were excluded as, at the time of this research, they did not provide an advanced search system based on the criteria defined in Section \ref{sec:methodology.search}. With a view  to mitigating  this threat, the search process was carried out twice since there was a delay of two months between the findings. Additionally, an exploratory search was conducted in the excluded databases along with the analysis of the returned papers.

\textbf{Internal validity.} Although this paper adopted a systematic mapping strategy, the inclusion and exclusion criteria were selected  on the basis of the authors' own  judgment. This means that some studies might  have been selected incorrectly. However, an attempt was made to attenuate this bias in  the systematic mapping protocol, by having an exhaustive discussion among the researchers to ensure there was  a common understanding. Moreover, this study followed open coding and axial coding procedures, which meant that the data extraction and findings were subject to a peer-review evaluation.

\textbf{External validity.} This paper performed a systematic mapping of MSR studies, regardless of the venue. This implies that the study may have overlooked some important  studies that did not match the data extraction process but were only published in MSR  venues. Thus, it is not possible to generalize the conclusions for all the MSR venues, such as MRSConf. However, the outcomes of this study allowed us to gain an insight into the importance of this topic from a wider range of relevant venues, e.g., ICSE, ESEM, and SANER.

\section{Conclusion}  \label{sec:conclusion}

MSR has grown into one of the key areas of Software Engineering research and is definitely a hot topic for new studies. The MSR protagonist has reflected on important venues that are only devoted to dealing with MSR research, and this has encouraged  researchers and practitioners to enhance the quality  and extend the range of the field. Given the importance of MSR, this study decided to update the research on MSR, first by carrying out a systematic mapping study to examine and throw light on the current state-of-the-art, and then  proposing an extended Cookbook that upgraded the recommendations, tools available, and number of lessons learned. The outcomes of 112 selected studies confirmed  that MSR is still growing, and showed there has been an increase of 84\% in the comments in this study when compared with the number in \textit{the original MSR Cookbook}. This study investigated the applied techniques of the selected studies, and found the most promising ones were Machine and Deep Learning. Future lines of research  should focus on keeping the Cookbook up–to-date, and enhancing  the mapping studies so that future trends can be understood, as well as demonstrating how MSR studies can lead to new recommendations and further discussion.

\begin{acks}
This work is supported by PROPG and DIRPPG from UTFPR, as well as National Council for Scientific and Technological Development (CNPq) (Grant no. 437937/2018-6).
\end{acks}



\end{document}